\documentclass[3p,times,twocolumn]{elsarticle}
\usepackage{url,ecrc}
\pdfoutput=1
\volume{00}
\firstpage{1}
\journalname{Nuclear Physics B Proceedings Supplement}
\runauth{}
\jid{nuphbp}
\jnltitlelogo{Nuclear Physics B Proceedings Supplement}
\usepackage{amssymb}
\usepackage{amsmath}
\usepackage{amsfonts}
\usepackage[figuresright]{rotating}
\begin{document}
\begin{frontmatter}
\dochead{}
\title{Global Bayesian Analysis of the Higgs-boson Couplings\tnoteref{fn}}
\tnotetext[fn]{Based on a talk presented by {\bf Diptimoy Ghosh} in the 
37$^{\rm th}$ International Conference on High Energy Physics (ICHEP) in Valencia from the 
2$^{\rm nd}$ to the 9$^{\rm th}$ July 2014.\\}
\author[a]{Jorge de Blas}
\ead{Jorge.DeBlasMateo@roma1.infn.it}
\author[b]{Marco Ciuchini}
\ead{marco.ciuchini@roma3.infn.it}
\author[a]{Enrico Franco}
\ead{enrico.franco@roma1.infn.it}
\author[a]{Diptimoy Ghosh}
\ead{diptimoy.ghosh@roma1.infn.it}
\author[c,d]{Satoshi Mishima}
\ead{satoshi.mishima@roma1.infn.it}
\author[e]{Maurizio Pierini}
\ead{maurizio.pierini@cern.ch}
\author[f]{Laura Reina}
\ead{reina@hep.fsu.edu}
\author[a]{Luca Silvestrini}
\ead{luca.silvestrini@roma1.infn.it}
\address[a]{INFN, Sezione di Roma, Piazzale A. Moro 2, I-00185 Roma, Italy}
\address[b]{INFN, Sezione di Roma Tre, Via della Vasca Navale 84, I-00146 Roma, Italy} 
\address[c]{Dipartimento di Fisica, Universit\`a di Roma ``La Sapienza'', 
Piazzale A. Moro 2, I-00185 Roma, Italy}
\address[d]{SISSA, Via Bonomea 265, I-34136 Trieste, Italy}
\address[e]{California Institute of Technology, Pasadena, California, USA}
\address[f]{Physics Department, Florida State University, Tallahassee, FL 32306-4350, USA}

\begin{abstract}
We present preliminary results of a bayesian fit to the Wilson coefficients of the 
Standard Model gauge invariant dimension-6 operators involving one or more Higgs fields, using data 
on electroweak precision observables and Higgs boson signal strengths. 
\end{abstract}
\end{frontmatter}
\section{Introduction}
\label{intro}

After a decades-long hunt, in the summer of 2012 the physics world erupted in excitement when both 
the ATLAS and CMS experiments at the Large Hadron Collider (LHC) at CERN announced their discovery 
of a particle that looked like the Higgs boson ($H$) \cite{Aad:2012tfa,Chatrchyan:2012ufa}. 
With the help of two-and-a-half times more data and sophisticated experimental analyses, it is now 
confirmed that the newfound particle behaves, indeed, very much like the Standard Model (SM) Higgs 
boson. That this Higgs boson decays to SM gauge bosons is now established with high statistical 
significance. In fact, each of the decay channels $H \to \gamma \gamma$, $H \to W^+ W^-$ and 
$H \to Z Z$ is by now a discovery channel. There is also good evidence of its 
non-universal couplings to fermions. The decays to $\tau^+ \, \tau^-$ and $b \, \bar{b}$ final 
states have also been seen with good confidence.  

Since the Higgs-boson mass ($m_H$) has now been measured, its couplings to SM particles are 
completely predicted except for the residual arbitrariness introduced by the Yukawa couplings to
fermions, which are nevertheless very constrained by the precise measurement of fermion masses.
This means that any deviation from the SM predictions will provide unambiguous evidence for New 
Physics (NP). Unfortunately, large deviations from the SM expectations are already ruled out (except 
possibly in the couplings to light fermions and/or $H \to Z \gamma$). This, in conjunction with the 
absence of any other direct NP signal so far, leads us to expect a deviation at the 
level of no more than a few percents. Hence, a rigorous study of the Higgs-boson couplings in the Run-II 
of the LHC and also in the high luminosity phase is mandatory. 

Although new particles at the TeV scale or below are perfectly allowed by the LHC data, it is interesting 
to study the sensitivity of the current Higgs-boson related 
measurements to short-distance physics assuming an effective field theory framework.      
The effect of heavy NP (beyond the reach of LHC for direct production) can be parametrized in 
terms of gauge-invariant higher-dimensional operators involving only SM fields.  
In this case, one supplements the SM Lagrangian with operators of mass dimension greater 
than 4, 
\begin{equation}
{\mathcal L}  = {\mathcal L}_{\rm SM} + \dfrac{1}{\Lambda} {\mathcal L}^{(5)} 
+ \dfrac{1}{\Lambda^2} {\mathcal L}^{(6)} + ..
\end{equation}
In the SM there is only one operator of dimension 5, the celebrated Weinberg operator 
which gives Majorana masses to light neutrinos \cite{Weinberg:1979sa}. 
As this operator is irrelevant for our discussion of Higgs physics, we will not consider it here. 
On the other hand, the number
of dimension-6 operators is much higher: even for one generation 
the count of the total number of operators grows to 59 \cite{Grzadkowski:2010es}
\footnote{The original work by Buchmuller and Wyler \cite{Buchmuller:1985jz} had 80 operators out 
of which only 59 were shown to be independent by the authors of \cite{Grzadkowski:2010es}.}. 
Adding general flavour structure increases this number to a gigantic 2499 \cite{Alonso:2013hga}.
For phenomenological explorations of some of these operators and related studies, see 
\cite{Han:2004az,delAguila:2011zs,
Carmi:2012in,
Biswal:2012mp,
Banerjee:2012xc,
Corbett:2012dm,
Corbett:2012ja,
Masso:2012eq,
Ghosh:2012ep,
Dumont:2013wma,
Einhorn:2013kja,
Einhorn:2013tja,
Boos:2013mqa,
Jenkins:2013zja,
Jenkins:2013wua,
Alonso:2013hga,
Brivio:2013pma,
Grojean:2013kd,
Contino:2013kra,
Contino:2014aaa,
Pomarol:2013zra,
Alloul:2013naa,
Ciuchini:2013pca,
Elias-Miro:2013gya,
Elias-Miro:2013mua,
Elias-Miro:2013eta,
Blas:2013ana,
Ellis:2014dva,
Belusca-Maito:2014dpa,
Biekoetter:2014jwa,
Beneke:2014sba,
Ghosh:2014wxa}.

In the following section we will choose one operator basis and introduce the set of operators 
considered in this work. The experimental data used in our analysis will be discussed in 
Sec.~\ref{data}. We will present our results in Sec.~\ref{results} and outline some conluding remaks 
in Sec.~\ref{conclusion}.

\section{Operator basis}
\label{op:basis}

Several operator bases have been used in the literature to describe the physics of 
gauge-invariant dimension-6 operators in the SM \cite{Hagiwara:1993ck,Giudice:2007fh,Grzadkowski:2010es}. 
In this work we concentrate on electroweak and Higgs-boson observables only. 
While depending on the set of observables chosen for a specific study one of these operator bases can 
be more convenient than others, physics should be basis independent. 
Moreover, we aim to study also other observables (e.g., flavour and other low-energy ones) in the near 
future. Therefore the choice of one operator basis is as good as any other one for our purpose.
As we do not want to introduce another new basis in the 
literature, we choose to adopt the fairly general basis introduced in Ref.~\cite{Grzadkowski:2010es}. 
As mentioned earlier, the total number of independent operators was shown to be 59. 
The basis of Ref.~\cite{Grzadkowski:2010es} consists of 15 bosonic operators, 
19 single-fermionic-current operators and 25 four-fermion operators for each 
fermion generation.
Since in this study we limit ourselves to electroweak and Higgs-boson signal 
strength observables (extending the previous work by some of us 
\cite{delAguila:2011zs,Ciuchini:2013pca,Blas:2013ana}), we consider 
only a subset 
of operators. In particular, we only consider operators involving one or more Higgs fields. 
Operators which involve fermionic fields are assumed to be flavour-diagonal and family 
universal. 
Moreover, we restrict this study to Charge-Parity ($CP$) even operators only. 
As the Wilson coefficients are generated at the scale $\Lambda$, ideally, one should also use the 
Renormalization Group Equations (RGE) to evolve them from the scale $\Lambda$ to the energy scale 
relevant for the process of interest\footnote{The anomalous dimension matrix for all the 2499 
operators has been computed recently in \cite{Jenkins:2013zja,Jenkins:2013wua,Alonso:2013hga}.}. 
In this work we neglect the effect of RGE. Below, we introduce our notations and list all the 
operators relevant for our study.  

\begin{itemize}
\item Bosonic operators: 
\begin{align}
\mathcal{O}_{HG} &= 
(H^\dagger H) \, G^A_{\mu\nu} G^{A\mu\nu}  \, \, , 
\\
\mathcal{O}_{HW} &= 
(H^\dagger H) \, W^I_{\mu\nu} W^{I\mu\nu} \, \, ,
\\
\mathcal{O}_{HB} &= 
(H^\dagger H) \, B_{\mu\nu} B^{\mu\nu} \, \, ,
\\
\mathcal{O}_{HWB} &= 
(H^\dagger\tau^I H) \, W^I_{\mu\nu} B^{\mu\nu} \, \, ,
\\
\mathcal{O}_{HD} &= 
(H^\dagger D^\mu H)^* \, (H^\dagger D_\mu H) \, \, ,
\end{align}
where $\tau^I$ are the three Pauli matrices.

The Wilson coefficients for the operators  $\mathcal{O}_{HWB}$ and $\mathcal{O}_{HD}$ 
(we denote them by $C_{HWB}$ and $C_{HD}$ respectively) are related to the well known 
Peskin and Takeuchi parameters $S$ and $T$ \cite{Peskin:1990zt} by,
\begin{align}
S &=  \dfrac{4 s_W c_W}{\alpha_{\rm em} (M_Z)} \dfrac{\mathrm{v}^2}{\Lambda^2} C_{HWB} \, \, , \\
T &=  - \dfrac{1}{2 \alpha_{\rm em} (M_Z)} \dfrac{\mathrm{v}^2}{\Lambda^2} C_{HD} \, \,, 
\end{align}
where $c_W$ and $s_W$ are the cosine and sine of the weak mixing angle $\theta_W$ respectively, 
$\mathrm{v}$ is the Vacuum Expectation Value (VEV) of the Higgs field and $\alpha_{\rm em}$ is the 
electromagnetic fine-structure constant.

In addition to the above operators, there are two more purely bosonic operators involving 
only the Higgs-boson field, namely, 
\begin{align}
\mathcal{O}_{H\Box} &=  (H^\dagger H) \Box (H^\dagger H) \, \, \rm and \\
\mathcal{O}_{H} &=  (H^\dagger H)^3  \, \, .
\end{align}
The operator $\mathcal{O}_{H\Box}$ contributes to the wave-function renormalization of the 
Higgs field and $\mathcal{O}_{H}$ contributes to the Higgs potential, i.e., the VEV $\mathrm{v}$ and 
the SM Higgs-boson self coupling $\lambda$. We will see later that this makes $\mathcal{O}_{H\Box}$ 
poorly constrained and $\mathcal{O}_{H}$, which does not affect our observables at all, 
remains unconstrained by our analysis. A joint measurement of the Higgs mass $m_H$ and the 
self-coupling $\lambda$ is required to constrain this operator. 
There are 8 more bosonic operators (6 $CP$ odd + 2 $CP$ even) in the total 15 bosonic operators 
listed in \cite{Grzadkowski:2010es}, but they either do not involve any Higgs field or are CP-odd. 
Thus, we do not consider them in our analysis.

\item Single-fermionic-current operators:
\begin{align}
\mathcal{O}_{HL}^{(1)} &= 
(H^\dagger i \overleftrightarrow{D}_\mu H) (\overline{L} \gamma^\mu L) \, \, ,
\\[1mm]
\mathcal{O}_{HL}^{(3)} &= 
(H^\dagger i \overleftrightarrow{D}_\mu^I H) (\overline{L}\, \tau^I \gamma^\mu L) \, \, ,
\\[1mm]
\mathcal{O}_{He} &= 
(H^\dagger i \overleftrightarrow{D}_\mu H) (\overline{e}_{R} \gamma^\mu e_{R}) \, \, ,
\\[1mm]
\mathcal{O}_{HQ}^{(1)} &= 
(H^\dagger i \overleftrightarrow{D}_\mu H) (\overline{Q} \gamma^\mu Q) \, \, ,
\\[1mm]
\mathcal{O}_{HQ}^{(3)} &= 
(H^\dagger i \overleftrightarrow{D}_\mu^I H) (\overline{Q}\, \tau^I \gamma^\mu Q) \, \, ,
\\[1mm]
\mathcal{O}_{Hu} &= 
(H^\dagger i \overleftrightarrow{D}_\mu H) (\overline{u}_{R} \gamma^\mu u_{R}) \, \, ,
\\[1mm]
\mathcal{O}_{Hd} &= 
(H^\dagger i \overleftrightarrow{D}_\mu H) (\overline{d}_{R} \gamma^\mu d_{R}) \, \, ,
\\[1mm]
\mathcal{O}_{Hud} &= 
i (\widetilde{H}^\dagger D_\mu H) (\overline{u}_{R} \gamma^\mu d_{R}) \, \, .
\end{align}
As we consider flavour diagonal couplings only, all the above operators except $\mathcal{O}_{Hud}$ 
are hermitian. 
Here, $\widetilde{H} = i \tau^2 H^*$ and the hermitian derivatives have been defined as, 
\begin{align}
H^\dagger \overleftrightarrow{D}_\mu H &= 
H^\dagger (D_\mu H) - (D_\mu H)^\dagger H \, \, \rm and 
\\
H^\dagger \overleftrightarrow{D}_\mu^I H &= 
H^\dagger \tau^I (D_\mu H) - (D_\mu H)^\dagger \tau^I H \, .
\end{align}

There are also (non-hermitian) operators involving scalar fermionic currents, 
\begin{align}
\mathcal{O}_{eH} &= 
(H^\dagger H) (\bar{L}\, e_{R} H) \, \, ,
\\
\mathcal{O}_{uH} &= 
(H^\dagger H) (\bar{Q}\, u_{R} \widetilde{H}) \, \, ,
\\
\mathcal{O}_{dH} &= 
(H^\dagger H) (\bar{Q}\, d_{R} H) \, \, .
\end{align}
Once the Higgs field gets a VEV, these operators modify the SM Yukawa couplings.  
There are 8 more operator structures which involve tensor fermionic currents. 
We do not consider them in the analysis presented here.   
\end{itemize}

\section{Experimental data}
\label{data}
In order to constrain the Wilson coefficients of the dimension-6 operators induced by NP, we use 
the data on (1) ElectroWeak Precision Observables (EWPO) from SLD, LEP-I, LEP-II and Tevatron and 
(2) Higgs signal strengths from ATLAS and CMS. The experimental values of the EWPO 
are summarized in Table~\ref{data-ew}.

\begin{table}[h]
\centering
\begin{tabular}{lc} 
\hline
$\alpha_s(M_Z^2)$                                         &  $0.1185\pm 0.0005$    \\
$\Delta\alpha_{\rm had}^{(5)}(M_Z^2)$                     &  $0.02750\pm 0.00033$  \\
$M_Z$ [GeV]                                               &  $91.1875\pm 0.0021$   \\
$m_t$ [GeV]                                               &  $173.34\pm 0.76$       \\
$m_H$ [GeV]                                               &  $125.5\pm 0.3$        \\
\hline
$M_W$ [GeV]                                               &  $80.385\pm 0.015$     \\
$\Gamma_W$ [GeV]                                          &  $2.085\pm 0.042$      \\
$\Gamma_{Z}$ [GeV]                                        &  $2.4952\pm 0.0023$    \\
$\sigma_{h}^{0}$ [nb]                                     &  $41.540\pm 0.037$     \\
$\sin^2\theta_{\rm eff}^{\rm lept}(Q_{\rm FB}^{\rm had})$ &  $0.2324\pm 0.0012$    \\
$P_\tau^{\rm pol}$                                        &  $0.1465\pm 0.0033$    \\
$\mathcal{A}_\ell$ (SLD)                                  &  $0.1513\pm 0.0021$    \\
$\mathcal{A}_{c}$                                         &  $0.670\pm 0.027$      \\
$\mathcal{A}_{b}$                                         &  $0.923\pm 0.020$      \\
$A_{\rm FB}^{0,\ell}$                                     &  $0.0171\pm 0.0010$    \\
$A_{\rm FB}^{0,c}$                                        &  $0.0707\pm 0.0035$    \\
$A_{\rm FB}^{0,b}$                                        &  $0.0992\pm 0.0016$    \\
$R^{0}_{\ell}$                                            &  $20.767\pm 0.025$     \\
$R^{0}_{c}$                                               &  $0.1721\pm 0.0030$    \\
$R^{0}_{b}$                                               &  $0.21629\pm 0.00066$  \\
\hline
\end{tabular}
\caption{Summary of experimental data on EWPO. \label{data-ew}}
\end{table}

\begin{table*}[ht!]
\center
\begin{tabular}{c|c|c|c}
               & {\bf Only EW}                & {\bf Only Higgs}             & {\bf EW + Higgs}             \\
\hline
               & $C_i/\Lambda^2$ [TeV$^{-2}$] & $C_i/\Lambda^2$ [TeV$^{-2}$] & $C_i/\Lambda^2$ [TeV$^{-2}$] \\
Coefficient    &            at 95\%           &          at 95\%             &          at 95\%             \\
\hline
$C_{HG}$       &   $--$                       & $[-0.0077,\, 0.0066]$        & $[-0.0077,\, 0.0066]$        \\
$C_{HW}$       & $--$                         & $[-0.039,\, 0.012]$          & $[-0.039,\, 0.012]$          \\
$C_{HB}$       & $--$                         & $[-0.011,\, 0.003]$          & $[-0.011,\, 0.003]$          \\
$C_{HWB}$      & $[-0.0082,\, 0.0030]$        & $[-0.006,\, 0.020]$          & $[-0.0063,\, 0.0039]$        \\
$C_{HD}$       & $[-0.025,\, 0.004]$          & $[-4.0,\, 1.4]$              & $[-0.025,\, 0.004]$          \\
$C_{H\Box}$    &                              & $[-1.2,\, 2.0]$              & $[-1.2,\, 2.0]$              \\
$C_{HL}^{(1)}$ & $[-0.005,\, 0.012]$          &         $--$                 & $[-0.005,\, 0.012]$          \\
$C_{HL}^{(3)}$ &  $[-0.010,\, 0.005]$         & $[-1.2,\, 0.3]$              & $[-0.010,\, 0.005]$          \\
$C_{He}$       & $[-0.015,\, 0.006]$          &        $--$                  & $[-0.015,\, 0.006]$          \\
$C_{HQ}^{(1)}$ &  $[-0.026,\, 0.041]$         & $[-28,\, 15]$                & $[-0.026,\, 0.041]$          \\
$C_{HQ}^{(3)}$ & $[-0.011,\, 0.013]$          & $[-0.6,\, 2.2]$              & $[-0.011,\, 0.013]$          \\
$C_{Hu}$       & $[-0.067,\, 0.077]$          & $[-5,\, 11]$                 & $[-0.067,\, 0.077]$          \\
$C_{Hd}$       & $[-0.14,\, 0.06]$            & $[-33,\, 15]$                & $[-0.14,\, 0.06]$            \\
$C_{Hud}$      &         $--$                 &       $--$                   &         $--$                 \\
$C_{eH}$       &         $--$                 & $[-0.071,\, 0.024]$          & $[-0.071,\, 0.024]$          \\
$C_{uH}$       &         $--$                 & $[-0.50,\, 0.59]$            & $[-0.50,\, 0.59]$            \\
$C_{dH}$       &         $--$                 & $[-0.073,\, 0.078]$          & $[-0.072,\, 0.078]$          \\
\hline
\end{tabular}
\caption{Fit results for the coefficients of the dimension six operators at 95\% probability. 
The fit is performed switching on one operator at a time.
Bounds from only EWPO, only Higgs signal strengths and the combined ones are shown separately. \label{res-1}}
\end{table*}

For the definitions and theoretical expressions of the EWPO and related issues, we refer the 
reader to \cite{Ciuchini:2013pca} and the references therein 
\footnote{For an update of their analysis see \cite{Satoshi-ichep}.}. The quantities in the first 
five rows in 
Table \ref{data-ew} have been used as inputs of our fit. Currently, we have used only their central 
values while fitting the NP coefficients. 

In addition to the EWPO, we also use the data on Higgs signal strengths from the ATLAS and CMS 
experiments. The theory prediction for the signal strength $\mu$ of one specific analysis
can be computed as,
\begin{equation}
\mu = \sum_i w_i  r_i  \, \,, 
\end{equation}
where the sum runs over all the channels which can contribute to the final state of the analysis.
The individual channel signal strength $r_i$ and the SM weight for that channel $w_i$ 
are defined as
\begin{align}
r_i &= \dfrac{[\sigma \times BR]_i}{[\sigma_{SM} \times BR_{SM}]_i} \, \, \rm and \\
w_i &= \dfrac{\epsilon_i [\sigma_{\rm SM} \times BR_{\rm SM}]_i}
{\sum_j \epsilon_j^{\rm SM} [\sigma_{\rm SM} \times BR_{\rm SM}]_j} \, .
\end{align}

In the presence of NP the relative experimental efficiencies, $\epsilon_i$, will in general be 
different from their values in the SM. In particular, the appearance of
new tensor structures 
in the vertices can modify the kinematic distribution of the final-state particles, thereby changing 
the efficiencies. In this work, we assume that this effect is negligible and use the SM weight 
factors throughout. This assumption is valid for small deviations from the SM couplings so that 
kinematic distributions are not changed significantly.

We have implemented our effective Lagrangian in FeynRules \cite{Alloul:2013bka} and used Madgraph 
\cite{Alwall:2014hca} to compute the NP contributions to the Higgs production cross sections 
numerically at the tree level. In order to compute the branching ratios we have used the formulae 
given in \cite{Contino:2014aaa} after changing them to our basis. We only consider NP effects 
which are linear (${\mathcal O} (1/\Lambda^2)$) in the dimension-6 operator coefficients.
In all cases, the SM $K$-factors\footnote{We define the SM $K$-factor to be the ratio of the cross section from the 
LHC Higgs Cross Section Working Group \cite{Heinemeyer:2013tqa} to the leading order number obtained using Madgraph.} 
have been used to estimate the effect of QCD corrections, even for the NP contributions. 
No theoretical uncertainties 
have been associated to the cross sections and branching ratios in our current analysis.

Experimental measurements for the signal strengths have been 
taken from Refs.~\cite{ATLAS-CONF-2013-012, CMS-PAS-HIG-13-001} for $H \to \gamma \gamma$, 
\cite{CMS-PAS-HIG-13-004,ATLAS-CONF-2013-108} for $H \to \tau \tau$, \cite{ATLAS-CONF-2013-030,Chatrchyan:2013iaa} for 
$H \to W^+  W^-$, and \cite{ATLAS-CONF-2013-013,CMS-PAS-HIG-13-002} for $H \to Z Z$.

\begin{table*}[ht!]
\center
\begin{tabular}{c|cc|cc|cc}
               &  \multicolumn{2}{c|}{{\bf Only EW}}         &  \multicolumn{2}{c|}{{\bf Only Higgs}} & \multicolumn{2}{c}{ {\bf EW + Higgs}} \\
\hline
               & \multicolumn{2}{c|}{$\Lambda$ [TeV]}        & \multicolumn{2}{c|}{$\Lambda$ [TeV]}   & \multicolumn{2}{c}{$\Lambda$ [TeV]}   \\
Coefficient    &        $C_i=-1$      &       $C_i=1$        &      $C_i=-1$     & $C_i=1$            & $C_i=-1$      & $C_i=1$       \\
\hline
$C_{HG}$       &  \hspace{-3mm}$--$   &  \hspace{-3mm}$--$   & $11.4$\ \ \ \     & $12.3$\ \ \        & $11.4$\ \ \ \ & $12.3$\ \ \   \\
$C_{HW}$       &  \hspace{-3mm}$--$   &  \hspace{-3mm}$--$   & $5.1$\ \ \ \      & $9.1$\ \ \         & $5.1$\ \ \ \  & $9.1$\ \ \    \\
$C_{HB}$       &  \hspace{-3mm}$--$   &  \hspace{-3mm}$--$   & $9.6$ \ \ \ \     & $17.2$\ \ \        & $9.6$ \ \ \ \ & $17.2$\ \ \   \\
$C_{HWB}$      &  $11.1$\ \ \ \       &  $18.4$\ \ \         & $12.5$\ \ \ \     & $7.1$\ \ \         & $12.6$\ \ \ \ & $15.9$\ \ \   \\
$C_{HD}$       &  $6.3$\ \ \ \        &  $15.4$\ \ \         & $0.5$\ \ \ \      & $0.8$\ \ \         & $6.3$\ \ \ \  & $15.5$\ \ \   \\
$C_{H\Box}$    &  \hspace{-3mm}$--$   & \hspace{-3mm}$--$    & $0.9$\ \ \ \      & $0.7$\ \ \         & $0.9$\ \ \ \  & $0.7$\ \ \    \\
$C_{HL}^{(1)}$ &  $14.8$\ \ \ \       & $9.2$\ \ \           & \hspace{-3mm}$--$ & \hspace{-3mm}$--$  & $14.8$\ \ \ \ & $9.2$\ \ \    \\
$C_{HL}^{(3)}$ &  $9.8$\ \ \ \        & $14.8$\ \ \          & $0.9$\ \ \ \      & $1.7$\ \ \         & $9.8$\ \ \ \  & $14.9$\ \ \   \\
$C_{He}$       &  $8.2$\ \ \ \        & $12.8$\ \ \          & \hspace{-3mm}$--$ & \hspace{-3mm}$--$  & $8.2$\ \ \ \  & $12.8$\ \ \   \\
$C_{HQ}^{(1)}$ &  $6.2$\ \ \ \        & $5.0$\ \ \           & $0.2$\ \ \ \      & $0.3$\ \ \         & $6.2$\ \ \ \  & $5.0$\ \ \    \\
$C_{HQ}^{(3)}$ &  $9.6$\ \ \ \        & $8.7$\ \ \           & $1.3$\ \ \ \      & $0.7$\ \ \         & $9.7$\ \ \ \  & $8.7$\ \ \    \\
$C_{Hu}$       &  $3.9$\ \ \ \        & $3.6$\ \ \           & $0.4$\ \ \ \      & $0.3$\ \ \         & $3.9$\ \ \ \  & $3.6$\ \ \    \\
$C_{Hd}$       &  $2.7$\ \ \ \        & $4.1$\ \ \           & $0.2$\ \ \ \      & $0.3$\ \ \         & $2.7$\ \ \ \  & $4.1$\ \ \    \\
$C_{Hud}$      & \hspace{-3mm}$--$    & \hspace{-3mm}$--$    & \hspace{-3mm}$--$ & \hspace{-3mm}$--$ & \hspace{-3mm}$--$  & \hspace{-3mm}$--$ \\
$C_{eH}$       & \hspace{-3mm}$--$    & \hspace{-3mm}$--$    & $3.8$\ \ \ \      & $6.4$\ \ \         & $3.8$\ \ \ \  & $6.4$\ \ \    \\
$C_{uH}$       & \hspace{-3mm}$--$    & \hspace{-3mm}$--$    & $1.4$\ \ \ \      & $1.3$\ \ \         & $1.4$\ \ \ \  & $1.3$\ \ \    \\
$C_{dH}$       & \hspace{-3mm}$--$    & \hspace{-3mm}$--$    & $3.7$\ \ \ \      & $3.6$\ \ \         & $3.7$\ \ \ \  & $3.6$\ \ \    \\
\hline
\end{tabular}
\caption{Lower bounds on the NP scale in TeV obtained by setting $C_i=\pm 1$. 
 \label{res-2}}
\end{table*}

\section{Results}
\label{results}
In our analysis we have used the Bayesian statistical approach. It has been implemented using the 
public package Bayesian Analysis Toolkit (BAT) \cite{2009CoPhC.180.2197C}. Flat priors have been 
chosen for the parameters to be fitted. 
We consider only one Wilson coefficient at a time and fit it first to the EWPO and Higgs-boson 
observables separately, and then to the combination of both.

Our results are summarized in Table \ref{res-1} where we show the 
95\% probability regions on the Wilson coefficients assuming the NP scale to be 1 TeV. It can be observed that except 
for ${\mathcal O}_{HWB}$ the Electroweak precision constraints are much stronger than the Higgs 
signal strength data for all the 
operators which contribute to the EWPO. The strong constraint on $C_{HWB}$ from the Higgs data 
is due to its contribution to the Higgs decay to two photons which is loop suppressed 
in the SM. More precisely, the direct NP contribution to the $H\gamma\gamma$ vertex can be written as, 
\begin{eqnarray}
{\mathcal L}_{NP} \subset \dfrac{\mathrm{v}}{\Lambda^2} \! \! \! \! & \! \!(- c_W s_W C_{HWB} + s_W^2 C_{HW} \nonumber \\ 
&+ c_W^2 C_{HB} )\, F_{\mu\nu}F^{\mu\nu}H \, ,
\label{hgaga}
\end{eqnarray}
which has to be compared with the SM vertex $c_\gamma\dfrac{\alpha_{\rm em}}{8 \pi \mathrm{v}}F_{\mu\nu}F^{\mu\nu}H$ 
with $c_\gamma \approx -6.48$. 
%
\begin{figure}[hb!]
\begin{center}
\begin{tabular}{c}
\includegraphics[scale=0.35]{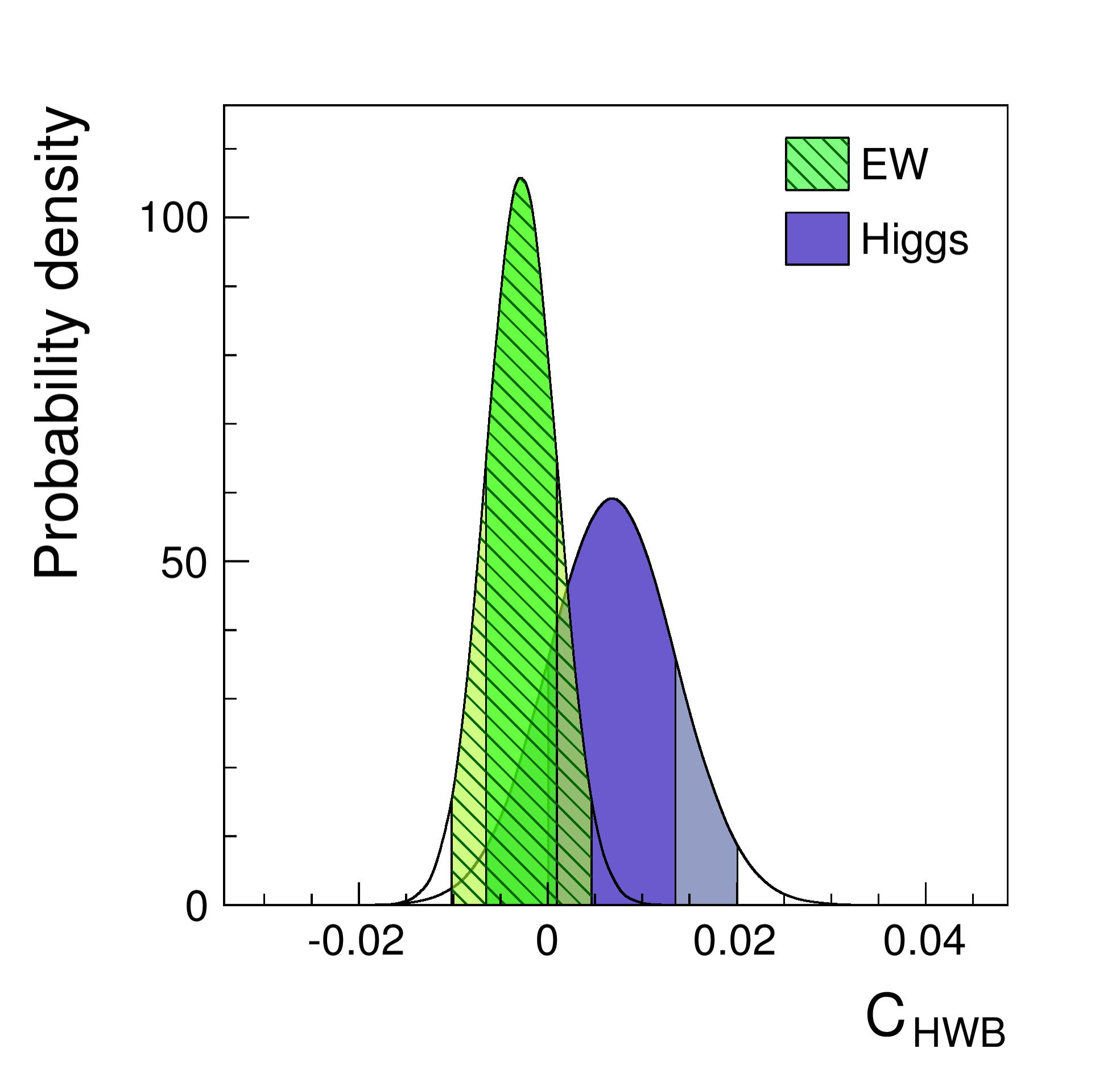} 
\end{tabular}
\caption{Posterior probabilities of $C_{HWB}$ considering only the EWPO (green) and 
only Higgs signal strengths (blue). The dark and light regions are 68\% and 95\% 
probability regions respectively. \label{fig:C_HWB}}
\end{center}
\end{figure}

In Fig.~\ref{fig:C_HWB} the posterior distribution of $C_{HWB}$ is shown with only EWPO and 
only Higgs signal strength data.  Eq.~(\ref{hgaga}) also explains why the bounds on $C_{HW}$ and 
$C_{HB}$ are rather strong from the Higgs signal strength data. 
The tight constraint on the operator ${\mathcal O}_{HG}$ can also be understood in a similar way. 
It contributes to the Higgs-boson production through gluon fusion, 
\begin{eqnarray}
{\mathcal L}_{NP} \subset \dfrac{\mathrm{v}}{\Lambda^2} C_{HG} \, G_{\mu\nu}^A G^{\mu\nu \, A}H \,,
\label{hgg}
\end{eqnarray}
which should be compared to the SM contribution 
$\dfrac{\alpha_s}{12 \pi \mathrm{v}}G_{\mu\nu}^A G^{\mu\nu \, A}H$, where $\alpha_s$ is the chromomagnetic 
fine-structure constant. 

The bounds on the dimension-6 operator coefficients in Table \ref{res-1} can also be translated into 
bounds on the NP scale for fixed values of the coefficients. 
We show them in Table \ref{res-2} for two values, $C_i = 1$ and $C_i = -1.$

A close look at the Table \ref{res-2} will reveal that, assuming $C_i(\Lambda) = \pm 1$, 
the lower bound on the NP scale for one of the operators that is constrained only by the Higgs 
data ($C_{H\Box}$) is less than 1 TeV. 
As this is close to the energy scale being probed at the LHC, the validity of such 
low bounds may be questionable. 

\section{Conclusion}
\label{conclusion}
The discovery of a Higgs boson and the absence of any other direct signal of new physics motivates 
to adopt effective field theories to study possible deviations of the Higgs-boson couplings from the SM. 
In this work we have taken the above route to study the effects of dimension-6 operators in 
Higgs physics. To this end, we have considered EWPO from LEP and Tevatron, and Higgs signal strength 
date from the LHC to fit the coefficients of the NP operators. In general, in an Ultraviolet (UV) 
complete model several operators are generated with specific relations among their 
coefficients. However, given the state of our knowledge about UV physics, any theoretical bias is 
premature and considering definite combinations of the operators in a fit is not strongly motivated. 
Here we have studied only one NP operator at a time. Barring accidental 
cancellations, our results should provide an estimate of the bounds even in relatively general scenarios. 
Updated results including more than one operator at a time 
will be presented in a future publication \cite{roma-mafias}.

The summary of our results is presented in Tables \ref{res-1} and \ref{res-2}. 
It is interesting that there is a strong hierarchy among the lower bounds on NP scales of different 
operators. It spans from cases with $\sim$ 1 TeV ($C_{H\Box}$) to \mbox{$\sim$ 15-20 TeV} (e.g., $C_{HB}$).

We observe that, except for the operator ${\mathcal O}_{HWB}$, the Higgs strength data is redundant 
for all the operators which contribute to the EWPO. 
The bound from Higgs data for the operator ${\mathcal O}_{HWB}$ is comparable to that obtained from 
EWPO. However, there are also operators 
(e.g., ${\mathcal O}_{HG, \, HW, \, HB }$) which are only constrained by the Higgs data. Moreover, as 
some of them contribute to loop-suppressed processes in the SM, the bounds on them are rather strong. 

To summarize, the preliminary results presented here indicate that the NP scale is beyond the reach 
of LHC energy for most of the operators 
if the Wilson coefficients are assumed to be $\pm 1$. However, these bounds can be weaker if the 
coefficients are smaller or 
specific correlations among the NP operators are present. Therefore NP scale of order $\sim$ TeV 
is allowed for perturbative values of the couplings.

\section*{Acknowledgments}
M.C. is associated with the Dipartimento di Matematica e Fisica, Universita di Roma Tre, 
and E.F. and L.S. are associated to the Dipartimento di Fisica, Universita di Roma ``La Sapienza''. 
The research leading to these results has received funding from the European Research Council
under the European Union’s Seventh Framework Programme (FP/2007-2013) / ERC Grant
Agreements n. 279972 ``NPFlavour'' and n. 267985 ``DaMeSyFla''. L.R. is also supported in part by 
the U.S. Department of Energy under grant DE-FG02-13ER41942.
\nocite{*}

\end{document}